\documentclass[twocolumn,showpacs,prb]{revtex4}

\usepackage{graphicx}
\usepackage{dcolumn}
\usepackage{bm}

\newcommand{\I}{{\rm i}}
\newcommand{\diff}{{\rm d}}

\begin{document}

\title{Density-matrix-renormalization-group study of excitons \\ in
  poly-diacetylene chains} 

\author{Gergely Barcza}
\author{\"Ors Legeza}
\affiliation{%
Research Institute for Solid-State Physics and Optics, Hungarian Academy of
Sciences, H-1121 Budapest, Hungary}
\author{Florian Gebhard}
\author{Reinhard M.\ Noack}
\affiliation{Department of Physics and Material Sciences Center, 
\\ Philipps-Universit\"at D-35032 Marburg, Germany}

\date{\today}

\begin{abstract}
We study the elementary excitations of a model Hamiltonian
for the $\pi$-electrons in poly-diacetylene chains.
In these materials, the bare band gap is only half the size of
the observed single-particle gap and the binding 
energy of the exciton of 0.5~eV amounts to 20\% of the single-particle
gap. Therefore, exchange and correlations 
due to the long-range Coulomb interaction require
a numerically exact treatment which we carry out using
the density-matrix renormalization group (DMRG) method.
Employing both the Hubbard--Ohno potential 
and the screened potential in one dimension,
we reproduce the experimental results for
the binding energy of the singlet exciton and
its polarizability. 
Our results indicate that there are optically dark states 
below the singlet exciton, in agreement with experiment. 
In addition, we find
a weakly bound second exciton with a binding energy of 0.1~eV.
The energies in the triplet sector do not match
the experimental data quantitatively, probably because
we do not include polaronic relaxation effects.
\end{abstract}


\maketitle

\section{Introduction}
\label{sec:intro}

\subsection{Experimental observations}
\subsubsection{Structure}
\label{subsec:structure}

Poly-diacetylenes (PDAs) are prototypical quasi one-dimensional 
materials.~\cite{PDAreview,Sariciftci}
Their monomer building unit is comprised of four carbon atoms.
The four outer electrons of each carbon atom are sp$^2$ hybridized.
Three of them form $\sigma$-bonds. The $\sigma$-bonds are between
neighboring carbon atoms on the chain and to covalent ligands 
$R$ and $R'$, which are several \AA\ long and differ for
various members of the PDA family.
The fourth electron is delocalized over the carbon backbone 
in a molecular $\pi$-orbital. 

\begin{figure}[htb]
\begin{center}
\includegraphics[height=2.4cm]{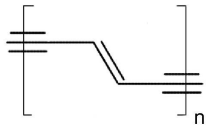}
\caption{Lewis structure of a poly-diacetylene unit cell.\label{Fig:structure}}
\end{center}
\end{figure}

The resulting Lewis structure is shown in Fig.~\ref{Fig:structure}.
The four carbon atoms in the unit cell are linked by a triple bond, 
a single bond, a double bond, and a single bond.
The atomic distances are $r_{\rm s}=1.4\, \mbox{\AA}$,
$r_{\rm d}=1.3\, \mbox{\AA}$, and $r_{\rm t}=1.2\, \mbox{\AA}$ 
for the single~(S), double~(D), and triple bonds~(T), respectively.
The chain of atoms is not perfectly straight; the single and double bonds
alternately form angles of $\varphi_1=120^{\circ}$ and $\varphi_2=240^{\circ}$ 
degrees. 

Very long undistorted polymer chains have been built 
starting from a monomer single crystal
so that the chains are perfectly ordered,~\cite{PDAreview}
and single polymer chains diluted in their monomer matrix
have even been prepared and studied.~\cite{Schottzucht}
In PDAs, exciton-polaritons have been generated that have been
shown to be coherent over tens of micrometers, i.e., several
ten thousand monomer units.~\cite{Schottexciton} 
Consequently, the opto-electronic properties of the PDAs 
result from the electrons' mutual interaction
and their interaction with the periodic lattice potential,
while the influence of disorder is negligible.

\subsubsection{Optical properties}
\label{subsec:opticalprops}

PDAs are insulators; the gap for single-particle excitation
is $\Delta\simeq 2.4\, {\rm eV}$. Band structure calculations
estimate the bare band gap to be $\Delta_{\rm bare}\approx 1.2\, {\rm eV}$,
i.e., electronic exchange and correlations account for half of the
single-particle gap. Moreover, the optical gap
for the primary exciton is $\Delta_{\rm opt}\simeq 1.9\, {\rm eV}$,
so that the exciton binding energy $\Delta_{\rm bind}\simeq 0.5\, {\rm eV}$
is about 20\% of the band gap. Due to the restricted geometry,
the electron-electron interaction must be treated accurately for the
calculation of the optical properties of the PDAs.

Lattice effects complicate the analysis of the spectra of PDAs in two ways.
First, the primary exciton with excitation energy
$\Delta_{\rm opt}$ is accompanied by phonon sidebands
that result from oscillations of the double and triple bonds.
These signals dominate the optical excitations below the band gap.
Second, some PDA single crystals 
such as TCDU~\cite{Japaner-redandbluechains}
exist in two different conformations, which have exciton energies
$\Delta_{\rm blue}=2.0\, {\rm eV}$ (`blue chains')
and $\Delta_{\rm red}=2.4\, {\rm eV}$ (`red chains'),
respectively. Therefore, it is not easy to disentangle
the effects of the electron-electron interaction
from those of the  electron-lattice interaction; the latter
gives rise to resonance shifts of several
tenths of an electron volt.

\subsection{Theoretical approaches}

\subsubsection{Extended Wannier theory}

In order to describe the optical excitations in PDAs,
two approaches have been taken. The first approach starts
from an ab-initio density-functional theory calculation
of the bare band structure in local-density approximation (LDA),
which is then supplemented by an approximate treatment of the residual
electron-electron interaction, e.g., the $G$$W$ approximation 
for the single-particle bands and the Bethe-Salpeter equation (BSE)
for the excitons (LDA+$G$$W$+BSE).~\cite{LouieRohlfing,LDA-GW-BSE}
Actual calculations for the PDAs often omit the $G$$W$~step 
(Wannier theory).~\cite{van-der-Horst}
Within this approach, a number of experimental data could be 
reproduced, e.g., the exciton binding energy
and its polarizability.
Unfortunately, the theory does not predict optically dark states
below the exciton resonance.

\subsubsection{Model calculations}

The second approach to a theoretical description
of the primary excitations in polymers starts from
a many-particle model Hamiltonian that describes
only the $\pi$-electrons and their mutual interaction.
The parameters for the kinetic energy of the electrons
are taken from LDA calculations, and the Coulomb interaction 
is approximated in various ways, e.g.,
with the Ohno parametrization~\cite{Ohno} of the Pariser--Parr--Pople 
potential.~\cite{PPP}
With the help of the DMRG
method,~\cite{steve} the ground state and elementary excitations
for such models can be calculated for large chains with a very high
accuracy. In this way, the electron-electron interaction
is treated without resorting to any approximations.
Unfortunately, specific calculations for the PDAs~\cite{Bursill}
have only had limited success in meeting the experimental test.

\subsubsection{Simplifying assumptions and outline of this work}

In this work, we consider structures which start with a triple bond and 
end with a double bond, i.e., we consider the sequences [(TSDS)$_{m-1}$TSD]
of $N=4m$ carbon atoms with $m$ triple bonds and $m$ double bonds.
We perform a DMRG study of 
a many-body model Hamiltonian for the $\pi$-electrons which uses 
Barford and Bursill's parametrization for the 
band-structure part~\cite{Bursill}
but employs the screened Coulomb potential in one dimension.
The essential difference between the screened potential
and the Ohno potential is that the local
interaction part is larger in the screened potential. Because of this, 
the Hubbard--Ohno potential leads to
a good description of the low-energy singlet excitations in PDAs.

We do not include lattice relaxation
for the single-particle and optical excitations. 
As has been shown by Barford and 
co-workers,~\cite{Barfordrelax} the relaxation energies for triplet
excitons in trans-polyacetylene
can be as large as $0.4\, {\rm eV}$. The parameter sets we will use below
apply to the rigid-lattice situation and 
will change slightly when lattice relaxation is taken into account
properly. 

The different varieties of PDA differ in their ligands, which introduce
local potentials on the carbon atoms with double bonds.
Therefore, the excitation energies of the PDAs differ by a few tenths
of an electron volt. In this work, we ignore the ligand effects
and only consider a prototypical case.

\section{Model for poly-diacetylene}
\label{sec:model}

\subsection{Operators in second quantization}
\subsubsection{Kinetic energy operator, current operator, and dipole operator}
\label{subsec:kinetic}

In this work we will 
restrict ourselves to the description of the $\pi$~electrons
because they dominate the optical response of the poly-diacetylenes
for energies $\hbar \omega<3\, {\rm eV}$.
The motion of the electrons is described by the operator for the
kinetic energy,
\begin{equation}
\hat{T} = - \sum_{l;\sigma} t_{l} 
\left( \hat{c}_{l,\sigma}^+\hat{c}_{l+1,\sigma}
+ \hat{c}_{l+1,\sigma}^+\hat{c}_{l,\sigma}\right) \; ,
\label{eqn:hatT}
\end{equation}
where $\hat{c}^+_{l,\sigma}$,
$\hat{c}_{l,\sigma}$ are creation and annihilation operators, respectively, for
electrons with spin~$\sigma=\uparrow,\downarrow$ on site~$l$
with three-dimensional coordinate $\vec{r}_{l}$.
The matrix elements
$t_{l}$ are the electron transfer amplitudes between neighboring
sites. Following Ref.~\onlinecite{Bursill} we set
\begin{equation}
t_{\rm s }=2.4494\, \mbox{eV} \quad, \quad
t_{\rm d }=2.7939\, \mbox{eV} \quad, \quad
t_{\rm t }=3.4346\, \mbox{eV}
\label{eq:bsparameters}
\end{equation}
for the single, double, and triple bonds, respectively.
We consider the half-filled band exclusively, i.e.,
the number of $\pi$~electrons~$N_{\rm e}$ equals the
number of lattice sites~$N$. 

The electrical current operator is given by
\begin{equation}
\hat{J}= -\I e a \sum_{l,m;\sigma} t_{l}\left( 
\hat{c}_{l+1,\sigma}^+\hat{c}_{l,\sigma} 
- \hat{c}_{l,\sigma}^+\hat{c}_{l+1,\sigma} \right) \; ,
\label{eq:current}
\end{equation}
where $a$ is an average bond length, i.e.,
we ignore geometry effects in $\hat{J}$ due to the difference in bond 
lengths.\cite{philmag1}

Finally, we define the operator for the dipole moment,
\begin{equation}
\hat{d}=\sum_l |\vec{r}_{l}-\vec{r}_1|\left(\hat{n}_l-1\right) \; .
\label{eqn:dipole}
\end{equation}
Here $\hat{n}_{l}=\hat{n}_{l,\uparrow}
+\hat{n}_{l,\downarrow}$ counts the number of electrons on site~$l$,
and $\hat{n}_{l,\sigma}= \hat{c}^+_{l,\sigma}\hat{c}_{l,\sigma}$ 
is the local density operator at site~$l$ for 
spin~$\sigma$. Recall that we treat the PDA chain
as perfectly straight, so that $|\vec{r}_{l}-\vec{r}_1|$
is the appropriate sum over the bond distances $r_{\rm s}$,
$r_{\rm d}$, and $r_{\rm t}$ for the single, double, and triplet
bonds between the sites~$l$ and~$m$.

\subsubsection{Coulomb interaction}
\label{subsec:Coulomb}

The electrons interact electrostatically via the Cou\-lomb interaction
(Pariser--Parr--Pople model~\cite{PPP})
\begin{eqnarray}
\hat{V} &=& \frac{U}{\epsilon_{\rm d}} 
\sum_{l} \left(\hat{n}_{l,\uparrow}-\frac{1}{2}\right)
\left(\hat{n}_{l,\downarrow}-\frac{1}{2}\right) \nonumber \\
&& + \frac{1}{2\epsilon_{\rm d}}\sum_{l\neq m} V(l-m)
\left[\left(\hat{n}_{l}-1\right)
\left(\hat{n}_{m}-1\right)\right] \; .
\label{eqn:ourinteraction}
\end{eqnarray}
The PDAs are insulators. Therefore, the Coulomb interaction
is not dynamically screened at the energy scale of a few electron volts,
and the screening is taken into account reasonably well
by a static dielectric screening with 
dielectric constant $\epsilon_{\rm d}=2.3$ for PDAs.

For the description of electrons and holes in quantum wires
and other quasi one-dimensional structures,
various effective potentials 
have been used in the literature.~\cite{Loudon,KochundCo}
For polymers, the general Pariser--Parr--Pople potential~\cite{PPP}
is often approximated by the semi-empirical 
Ohno potential,~\cite{Ohno,Bursill,Baeriswylbook}
\begin{equation}
V^{\rm Ohno}(l-m) = 
\frac{V}{\sqrt{1+\beta (|\vec{r}_{l}-\vec{r}_{m}|/\mbox{\AA})^2}}
\label{eqn:Ohonodef}
\end{equation}
with $U=V$.
At large distances, the Coulomb interaction must be recovered.
Therefore, we require
\begin{equation}
V(l-m) \approx 
\frac{e^2}{|\vec{r}_{l}-\vec{r}_{m}|} \quad \mbox{for}
\quad |\vec{r}_{l}-\vec{r}_{m}| \gg \frac{\mbox{\AA}}{\sqrt{\beta}} \; ,
\end{equation}
which implies
$\sqrt{\beta}=V/(14.397\, {\rm eV})$, 
where we have used that $e^2/(2a_{\rm B})=13.605\, {\rm eV}$ is the
Rydberg energy
and $a_{\rm B}=0.5291\, \mbox{\AA}$ is the Bohr radius. 
The remaining free parameter~$V$ describes the modification
of the Coulomb potential at short distances due to
the confinement of the electrons to the chain.
Below, we derive the Ohno potential and justify it
for intermediate to large length scales. For short distances, however,
the Hubbard interaction must be kept explicitly, leading to
an additional parameter.

\subsubsection{Effective Coulomb potentials}
\label{subsec:Ohnoderived}

Our derivation of the one-dimensional effective potentials for the various cases
closely follows Refs.~\onlinecite{Kochbuch} and \onlinecite{Hoyerdiss}.
In order to set up the single-particle basis in which the
one-dimensional Hamiltonian~(\ref{eqn:ourinteraction}) is formulated,
we solve the single-particle Schr\"odinger equation
for electrons whose motion is restricted to the $z$-direction
due to a confining potential. We set $W_{\rm conf}(x,y,z)=W_2(x,y)W_1(z)$,
so that the single-particle wave functions factorize:
$\Psi(x,y,z)=\xi(x,y)\phi(z)$. The potential $W_1(z)$ incorporates the
(small) effects of the various ligands and permits a discrimination
of the poly-diacetylenes. For the purpose of this work, 
we set it to a constant which we absorb into $W_2(x,y)$.

The confining potential perpendicular
to the chain direction is assumed to be very strong so that 
the electron wave function in the direction perpendicular to the chain
is given by the lowest-energy state~$\xi_0(x,y)$.
The effective Coulomb potential between
two charges at distance~$|z|$ is then given by~\cite{Kochbuch,Hoyerdiss}
\begin{equation}
V^{\rm eff}(z) = 
\int \diff x \diff y \diff x' \diff y' 
\frac{e^2|\xi_0(x,y)|^2 |\xi_0(x',y')|^2 }{
\sqrt{(x-x')^2+(y-y')^2+z^2}} \; . \label{app:Veff}
\end{equation}
This expression can be simplified further for a parabolic 
confining potential,~\cite{Hoyerdiss}
\begin{equation}
W_2(x,y)= \frac{1}{2} m \omega_{\rm conf}^2 (x^2+y^2)
=\frac{2\hbar^2 (x^2+y^2)}{m R^4} \; ,
\end{equation}
where the parameter~$\omega_{\rm conf}=2\hbar/(mR^2)$ 
characterizes the strength of the confining 
potential. The ground-state wave function of the harmonic oscillators
in the $x$ and $y$ directions is given by
\begin{equation}
\xi_0(x,y) = \sqrt{\frac{2}{\pi}} 
\frac{1}{R} \exp\left(-\frac{x^2+y^2}{R^2}\right) \; ,
\label{eqn:xiwf}
\end{equation}
where $|\xi_0(x,y)|^2$ is a Gaussian with standard deviation
$\Delta x = \Delta y=R/2$. This means that we find the electrons
in the region $(|x|\leq R, |y|\leq R)$ with a probability of more
than 90~percent. In poly-diacetylene single crystals,~\cite{Weiser} 
the distance between chains is typically $d \simeq 5\, \mbox{\AA}$ so that
the condition of weak overlap between the chains,
$R\lesssim d$, is fulfilled for $R=3.6\, \mbox{\AA}$.
This view is supported by the fact that
the optical excitations for poly-diacetylene single crystals 
do not differ much from those for single chains diluted in
their single-crystal monomer matrix.~\cite{HorvathWeiserLapersonne}
The excitation energy to the next confinement level is $\hbar\omega_{\rm conf}
=15.235\, {\rm eV}/(R/\mbox{\AA})^2$. 
For $R=3.6\, \mbox{\AA}$, the excitation energy to the next confinement level
is $\hbar\omega_{\rm conf}=1.2\, {\rm eV}$ so that higher confinement
levels are not thermally populated at room temperature.

When we insert~(\ref{eqn:xiwf})
into~(\ref{app:Veff}), we can carry out the Gaussian integrals in 
the coordinates $X=(x+x')/2$ and $Y=(y+y')/2$ and are left with
a double integral over $x^r=x-x'$ and $y^r=y-y'$.
In polar coordinates, the resulting angular integral becomes trivial 
and we find~\cite{Hoyerdiss}
\begin{eqnarray}
V^{\rm scr}(z) &=& \frac{e^2}{R^2} 
\int_{0}^{\infty} \diff r (2r) 
\frac{\exp\left(-(r/R)^2\right)}{\sqrt{r^2+z^2}} \nonumber \\
&=& \frac{e^2}{R} \sqrt{\pi}
\exp\left[\left(z/R\right)^2\right] 
\left[1-{\rm erf}\left(|z|/R\right)\right] \, , 
\label{appeq:result}
\end{eqnarray}
where ${\rm erf}(x)$ is the error function.

\begin{figure}[htb]
\begin{center}
\includegraphics[width=8cm]{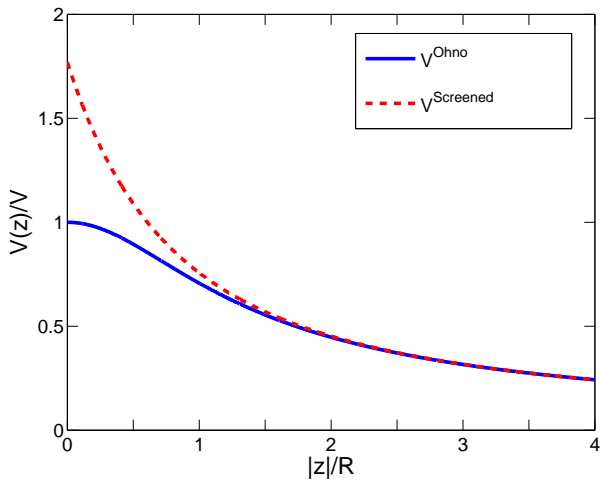}
\caption{(Color online) Ohno potential $V^{\rm Ohno}(z)/V$ 
from Eq.~(\protect\ref{eqn:Ohnoforfigure}) (full line) and screened potential 
$V^{\rm scr}(z)/V$ from Eq.~(\protect\ref{eqn:Veffforfigure}) (dashed line)
as a function of $|z|/R$.\label{Fig:OhnovseffectiveV}}
\end{center}
\end{figure}

We compare the long-distance limit of~(\ref{appeq:result}) with
that of the Ohno potential~(\ref{eqn:Ohonodef}). We demand 
the coefficients of the order
$1/|z|$ and $1/|z|^3$ agree and, with the help 
of Eq.~(7.1.23) in Ref.~\onlinecite{Abramovitz}, find that
\begin{equation}
V=\frac{e^2}{R}=\frac{14.397\, {\rm eV}}{(R/a_0)} \, ,
\label{eq:VandR}
\end{equation}
or $R/a_0=1/\sqrt{\beta}$, with $a_0 = 1\, \mbox{\AA}$ the unit of length.
In terms of the confinement parameter~$R$, the screened potential
and the Ohno potential can be cast into the form
\begin{eqnarray}
V^{\rm Ohno}(z) &=& \frac{V}{\sqrt{1+(z/R)^2}} \; ,
\label{eqn:Ohnoforfigure} \\
V^{\rm scr}(z) &=& \sqrt{\pi} V \exp\bigl[\left(z/R\right)^2\bigr] 
\left[1-{\rm erf}\left(|z|/R\right)\right] ,
\label{eqn:Veffforfigure}
\end{eqnarray}
whose large-distance expansions differ only to order $(R/|z|)^5$.
A comparison of the screened potential and the Ohno potential for all
distances is shown in Fig.~\ref{Fig:OhnovseffectiveV}.
The agreement is to better than 10\% for all $|z|>R$. 
Even at $|z|=R/2\approx 1.8\, \mbox{\AA}$, the discrepancy is only 25\%.
Therefore, it is justified to 
replace the screened potential by the Ohno potential for
intermediate distances.
At short distances, the differences between the Ohno potential
and the screened potential are substantial.
The Ohno potential at $z=0$ is 
$V^{\rm Ohno}(0)=V$, in contrast to $V^{\rm scr}(z=0)=\sqrt{\pi}V$ 
with $\sqrt{\pi}\approx 1.77$. 

The above derivation applies to a straight geometry,
not to the zigzag geometry of the Lewis structure in poly-diacetylenes.
We account for the corresponding reduction of the chain length 
by the approximation $|z|\approx|\vec{r}_{l}-\vec{r}_{m}|$
as in the Ohno potential~(\ref{eqn:Ohonodef}), thereby ignoring the minor
changes due to the non-orthogonality of the chain axis and the $x$-axis.

\subsection{Hamilton operator}
\subsubsection{Hubbard--Ohno potential and screened potential}

In this work, we study the correlated motion
of electrons along a chain, which we model using the Hamiltonian
\begin{equation}
\hat{H}=\hat{T} + \hat{V} \label{eqn:hamilt} \, ,
\end{equation}
with $\hat{T}$ given by Eq.~(\ref{eqn:hatT}) and 
the potential $\hat{V}$ by Eq.~(\ref{eqn:ourinteraction}).

We use the Hubbard--Ohno potential
\begin{eqnarray}
V(l-m) &= &
\frac{V_{\rm HO}}{\sqrt{1+(|\vec{r}_{l}-\vec{r}_{m}|/R_{\rm HO})^2}} 
\label{eq:ohno-hubbard}
\end{eqnarray}
for $l\neq m$ and  $U_{\rm HO}=\sqrt{\pi} V_{\rm HO}$.
Here the effective Coulomb parameter~$V_{\rm HO}$ is linked to the
screening length~$R_{\rm HO}$ via Eq.~(\ref{eq:VandR}), 
$V_{\rm HO}=14.397\, {\rm eV}/(R_{\rm HO}/a_0)$.
For example, we find $R_{\rm HO}=3.6\, \mbox{\AA}$ 
for $V_{\rm HO}=4.0\, {\rm eV}$.
The Hubbard--Ohno potential was used earlier by 
Chandross, Mazumdar {\it et al.}~\cite{ChandrossMazumdar} to
explain the main absorption features of poly(para-phenylene vinylene) (PPV)
with the parameter set $V_{\rm CM}=4.0\, {\rm eV}$ and
$U_{\rm CM}=8.0\, {\rm eV}\approx \sqrt{\pi}V_{\rm CM}$. 
Our derivation in Sect.~\ref{subsec:Ohnoderived} justifies their 
modification of the Ohno potential, and their results for PPV support
our choice for $V_{\rm HO}=4.0\, {\rm eV}$.

For comparison, we also give results for the screened potential,
\begin{eqnarray}
V^{\rm scr}(l-m) &=& 
V_{\rm scr} \sqrt{\pi}
\exp\left[\left(\frac{|\vec{r}_{l}-\vec{r}_{m}|}{R_{\rm scr}}\right)^2\right]
\nonumber \\
&& \times \left[1-{\rm erf}\left(\frac{|\vec{r}_{l}-\vec{r}_{m}|}{R_{\rm scr}}
\right)\right] 
\; ,
\label{eq:erf}
\end{eqnarray}
which implies $U_{\rm scr}=\sqrt{\pi}V_{\rm scr}$. Again, we have
$V_{\rm scr}=14.397\, {\rm eV}/(R_{\rm scr}/\mbox{\AA})$.
For example, we find $R_{\rm scr}=4.1\, \mbox{\AA}$ 
for $V_{\rm scr}=3.5\, {\rm eV}$.

\subsubsection{Particle-hole symmetry}
\label{subsec:phtransform}

The Hamiltonian~(\ref{eqn:hamilt}) and the current operator~(\ref{eq:current})
are invariant under the particle-hole transformation 
$\hat{c}_{l,\sigma}\mapsto (-1)^l\hat{c}_{l,\sigma}^+$. 
At half band-filling, 
the ground state $|\Phi_0\rangle$
is also invariant under this transformation. 
Therefore,
the expectation value of the dipole operator~(\ref{eqn:dipole})
vanishes in the ground state,
$d_0=\langle \Phi_0 |\hat{d}| \Phi_0\rangle=-d_0=0$.
Likewise, the expectation values of the current operator vanishes
in the ground state, $\langle \Phi_0 | \hat{J} | \Phi_0\rangle=0$.

We expect the same relation for excitons (bound particle-hole excitations
of the ground state at half band-filling).
Therefore, in the presence of a weak electrical field,
such states show a quadratic Stark effect, see Sect.~\ref{subsec:Starkeffect}.

\section{Method}
\label{sec:method}

\subsection{Single-particle gap and exciton binding energy}
\label{subsec:energies}

Our model description contains a single parameter that can be taken to be
the screening length~$R$ or the Coulomb parameter~$V$.
We use this parameter to adjust the typical band gap in poly-diacetylenes.
The band gap or single-particle gap is defined by the difference
in chemical potentials for a system with $N_{\rm e}$ and $N_{\rm e}-1$ particles,
\begin{eqnarray}
\Delta &=& \mu(N_{\rm e})-\mu(N_{\rm e}-1)\\
&=& \left[E_0(N_{\rm e}+1)-E_0(N_{\rm e})\right]
-\left[E_0(N_{\rm e})-E_0(N_{\rm e}-1)\right] \; ,\nonumber
\end{eqnarray}
which, due to particle-hole symmetry at half band-filling, reduces to
\begin{equation}
\Delta = 2\left[E_0(N_{\rm e}=N+1)-E_0(N_{\rm e}=N)\right] \; ,
\end{equation}
where $E_0(N_{\rm e})$ is the energy of the $N_{\rm e}$-particle ground 
state~$|\Phi_0\rangle$.
Optical excitations to above the
single-particle gap are extended and thus
can transport current through the system.
Experimentally, they can be monitored by the onset of
the Franz--Keldysh oscillations in the electro-absorption signal.
Typical values for poly-diacetylene single crystals are
$\Delta = 2.3\, {\rm eV}$ for DCHD and $\Delta = 2.5\, {\rm eV}$ for 
PTS and PFBS.~\cite{Weiser}

In poly-diacetylenes, the singlet exciton and its vibronic replicas
carry most of the oscillator strength of the optical excitations.
The quadratic Stark effect in the electro-absorption proves that
they are bound states of electron-hole excitations.~\cite{Weiser}
The exciton energy thus defines the optical gap,
\begin{equation}
\Delta_{\rm opt} = E_{\rm ex}(N_{\rm e}=N)-E_0(N_{\rm e}=N) \;, 
\end{equation}
where $E_{\rm ex}(N_{\rm e}=N)$ is the energy of the first excited state of the
half-filled system~$|\Phi_{\rm ex}\rangle$,
which has a finite overlap with an optical excitation of the ground state,
$\hat{J}|\Phi_0\rangle$.
The binding energy of the excitons is then obtained as
\begin{equation}
\Delta_{\rm bind} = \Delta - \Delta_{\rm opt}\; .
\end{equation}
For DCHD and PTS/PFBS PDA 
single crystals,
the corresponding binding energies are
$\Delta_{\rm bind} = 0.48\, {\rm eV}$ and 
$\Delta_{\rm bind} = 0.51\, {\rm eV}$, respectively.~\cite{Weiser}

In our numerical investigation, we calculate the ground state $|\Phi_0\rangle$
and excited states $|\Phi_s\rangle$, $s\geq 1$.
An optical excitation with the
current operator $\hat{J}$, Eq.~(\ref{eq:current}), 
has the oscillator strength
\begin{equation}
w_s = \frac{|\langle\Phi_s|\hat{J}|\Phi_0\rangle|^2}%
{\langle \Phi_0|\hat{J}^2|\Phi_0\rangle}\leq 1 \quad ; 
\quad \sum_{s\geq 1} w_s=1 \; .
\label{eq:weights}
\end{equation}
Amongst the excited states $|\Phi_s\rangle$, 
we identify the exciton state $|\Phi_{\rm ex}\rangle$
as the energetically lowest-lying excitation that carries significant 
optical weight, $w_{\rm ex}> 0.1$. 
The same description can be obtained using the dipole operator 
given by Eq.~(\ref{eqn:dipole}).

\subsection{Polarizability and exciton wave function}
\label{subsec:Starkeffect}

Since it is a bound state, the exciton displays a quadratic Stark effect, i.e.,
the redshift of the resonance level with an external static electrical
field of strength~$F$ is given by
\begin{equation}
\delta \Delta_{\rm opt} = -\frac{1}{2} p F^2 \; ,
\label{eq:Starkshift}
\end{equation}
where $p$ is the polarizability. Note that the experiment
measures the Stark shift both of the ground state~$|\Phi_0\rangle$
and of the exciton~$|\Phi_{\rm ex}\rangle$.
In our calculations, we determine $E_0(F)$ and $E_{\rm ex}(F)$ 
for various fields~$F$ from the Hamiltonian 
\begin{equation}
\hat{H}(F) = \hat{H} - e F \hat{d} 
\label{eqn:hamiltwithF}
\end{equation}
with the dipole operator $\hat{d}$, see Eq.~(\ref{eqn:dipole}).
Note that we determine the
polarizability as measured in experiment, i.e., we need not resort to
further theoretical considerations here.

In order to extract the `exciton radius' from the polarizability,
one can start from the Frenkel picture~\cite{Tokura} or from the Wannier 
picture.~\cite{Weiser} As demonstrated in Ref.~\onlinecite{analysis},
the probability distribution $P_{\rm ex}(l,m)$
provides a very detailed picture of the spatial character of the exciton
in a many-particle approach.
It describes the particle-hole content of
the exciton wave function with respect to the ground state, i.e.,
it gives the probability that $|\Phi_{\rm ex}\rangle$
is an electron-hole excitation of the ground state~$|\Phi_0\rangle$
at sites~$l$ and~$m$, respectively. Explicitly,
\begin{eqnarray}
P_{\rm ex}(l,m)&=&\frac{p_{\rm ex}(l,m)}{\sum_{l,m}p_{\rm ex}(l,m)}\label{pij}\\
\noalign{\noindent and}\nonumber\\[-12pt]
p_{\rm ex}(l,m)&=& \sum_{\sigma}\left|\left\langle\Phi_{\rm ex}\right|
\hat{c}^{+}_{l,\sigma}\hat{c}^{}_{m,\sigma}
\left|\Phi_0\right\rangle\right|^2 \label{Pij} \; .
\end{eqnarray}
We denote the probability density to find an electron-hole pair 
at a separation~$r_{\rm eh}$ by $\overline{P}_{\rm ex}(r_{\rm eh})$,
\begin{equation}
\overline{P}_{\rm ex}(r) 
= \sum_{l,m} P_{\rm ex}(l,m)
\delta\left(r-|\vec{r}_l-\vec{r}_m|\right) \; .\label{pbar}
\end{equation}
The average electron-hole distance~$r_{\rm eh}$ 
is then obtained from
\begin{equation}
r_{\rm eh} = \langle r\rangle_{\rm ex} =
\int \diff r\,  r \overline{P}_{\rm ex}(r)
=  \sum_{l,m} P_{\rm ex}(l,m) |\vec{r}_l-\vec{r}_m| \; .
\label{<r>} 
\end{equation}
In Ref.~\onlinecite{Weiser}, a simple two-level model was considered
in which the exciton couples to a (representative) continuum
state. As we shall see in Sect.~\ref{sect:polari}, the two corresponding
electron-hole distances compare well with each other.

\subsection{Numerical procedure}
\label{Subsec:DMRG}

We present results both for the Hubbard--Ohno potential,
Eq.~(\ref{eq:ohno-hubbard}),
and for the screened potential, Eq.~(\ref{eq:erf}).
In the presence of long-ranged Coulomb interactions, 
a high numerical accuracy is of crucial importance. 
Therefore, we devote this subsection to the problem of how 
we determine and control the accuracy of our calculations.

In this work, we have performed the numerical calculations on finite chains
with open boundary condition (OBC) 
using the non-local version~\cite{xiang,nishimoto,legeza-dbss}
of the DMRG technique.\cite{white} 
The number of block states has been selected according 
to the dynamic block-state selection (DBSS)
approach.\cite{legeza-dbss,legeza-qdc} 

\subsubsection{Ground state and single-particle gaps}

In order to calculate the band and spin gaps, 
we have determined the lowest-lying eigenstates 
of various spin and charge sectors 
from independent DMRG runs.
In this case, the DBSS approach~\cite{legeza-dbss,legeza-qdc} 
permits a rigorous control of the numerical accuracy
because we can fix the threshold value of the quantum information loss~$\chi$. 
Here we take $\chi=10^{-5}$.
As another check, we have used the entropy sum rule 
for finite chain lengths for each DMRG sweep, i.e., we have verified 
that the sum rule has been satisfied after the third sweep. 
During our calculations, the maximum number of block states was varied 
in the range $256<M_{\rm max}<400$ for OBC due to the
large spin and band gaps. For $M_{\rm max}=400$, the
maximal chain length is $N_{\rm max}=150$. 
For these parameters, the individual states can be treated reliably 
with an accuracy given by $\chi=10^{-5}$.

In the presence of long-ranged interactions it is crucial 
to use a large $M_{\rm min}$ in order to provide a good
environment block, i.e., to maximize the Kullback--Leibler 
entropy.~\cite{legeza-entropy} Here we take $M_{\rm min}=128$. 
Therefore, we have kept
a number of block states that have small weight during 
the system build-up in the infinite-lattice step. During
the sweeping iterations of the finite-lattice part of the DMRG, 
they gain more weight,~\cite{legeza-entropy} so that they
subsequently become important for an accurate description of the ground state
and the excited states.  

\subsubsection{Optical excitations}

In order to calculate the optical gap, 
we have simultaneously calculated $N_{\rm s}$ low-lying eigenstates 
of the half-filled charge sector.
In this case, it is necessary to specify how
the block entropy is calculated from the target states.
In the presence of several target states, it is possible to derive
an upper and a lower bound for 
the mutual information between the system block and the environment block.
Therefore, an upper bound~\cite{kholevo1} 
and a lower bound~\cite{jozsa4} can be
derived for the accessible information,~\cite{legeza-qdc} but an exact
expression is not available. 

In our work, we have defined the reduced density matrix before truncation  
as $\rho=\sum_a p_a \rho_a$, where the $\rho_a$ are the reduced density matrices
for the individual target states, and we have used fixed weights
$p_a>0$, $\sum_a p_a = 1$.
We have tried various sets of values for $p_a$ in order to stabilize 
the calculations and to improve the accuracy. 
For most of the results presented here, we have found 
that the statistically independent choice, $p_a=1/N_{\rm s}$,
provides satisfactory results.
We have set the minimum number of block states to $M_{\rm min}=400$,
and the maximal number of block states used in our calculations 
is $M_{\rm max}=800$. 
We note in passing that the Davidson diagonalization routine gives 
stable results only for $N_{\rm s}\leq 2$.
For all cases of interest, $N_{\rm s}>2$, we have used
the Lanczos method in order to obtain stable results.

In order to identify the exciton state, 
we begin our calculations with $10\leq N_{\rm s}\leq 15$ target states
with a reduced demand in accuracy.
Once we have found the dominant optical excitation from the 
oscillator strengths~(\ref{eq:weights}), we 
repeat our calculations using the smaller 
number of target states actually required, typically $N_{\rm s}=5$.
We independently determine the exciton state from the optical weights
based on the current operator, Eq.~(\ref{eq:current}), and 
the electrical dipole operator, Eq.~(\ref{eqn:dipole}).
 
\subsubsection{Chain topology}

In the standard DMRG procedure for OBC, it is more efficient 
to treat models which possess reflection symmetry;
computational costs can be reduced significantly by applying the
symmetry.
In turn, the results are usually more accurate 
for the same parameter set and computer resources
as compared to a non-reflection-symmetric configuration.
In our study, a reflection-symmetric configuration can be realized 
by an appropriate choice of the bond sequence along the chain, e.g., 
[T(SDST)$_m$] with $N=4m+2$ carbon atoms
or [(SDST)$_{m-1}$SDS] with $N=4m$ carbon atoms.
This approach, however, has a drawback. We have found
that, for the reflection-symmetric configurations, end excitations 
that are similar to the end spins 
for the $S=1$ Heisenberg chain with OBC appear.~\cite{white-endspin}
Due to these extra degrees of freedom, the ground state 
becomes four-fold degenerate in the thermodynamic 
limit so that numerical calculations become less stable, 
especially when several target states are used to calculate the optical gap. 
In order to remove such end excitations, we could have modified 
the first and last spins or their couplings, as was done
in Ref.~\onlinecite{white-endspin}, or could have used the reflection
symmetry as a conserved quantum number.
In this work, we avoid these complications by using a chain configuration 
that does not have inversion symmetry, i.e., we use the bond sequence
[(TSDS)$_{m-1}$TSD] with $N=4m$ carbon atoms.

In general, we have calculated
the low-lying energy spectrum for both the symmetric 
and the non-symmetric chain configurations
in order to identify the excitations unambiguously.
In addition, we have determined the optical gap
and the dipole matrix elements
for the current and dipole operators for both types of chain configurations.   
In the remainder of this paper, we present our results
for the configuration [(TSDS)$_{m-1}$TSD].

\subsubsection{Finite-size scaling}

The PDAs are charge and spin insulators, i.e., the gaps for single-particle, 
optical, and magnetic excitations are finite. The materials are characterized
by finite correlation lengths. Therefore, end effects decay 
exponentially, and local operators that are calculated in the middle
of the chain display a regular behavior as a function of inverse system size.
Thus, various quantities that we calculate for finite chain lengths~$N$
can be extrapolated reliably to the thermodynamic limit, $N\to\infty$, 
by using a second-order polynomial fit. 

\begin{figure}[hb]
\begin{center}
\includegraphics[width=7.9cm]{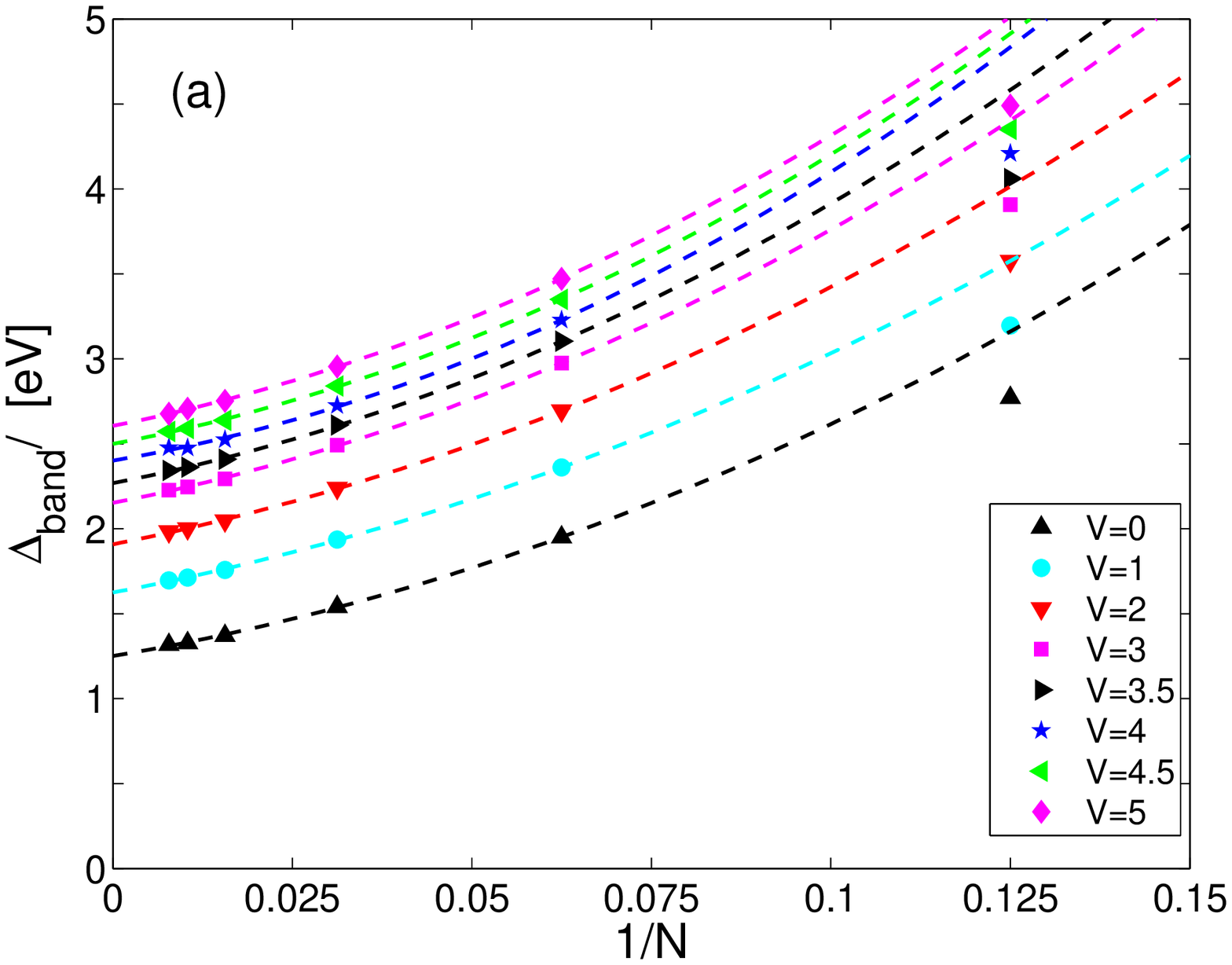} \\
\includegraphics[width=7.9cm]{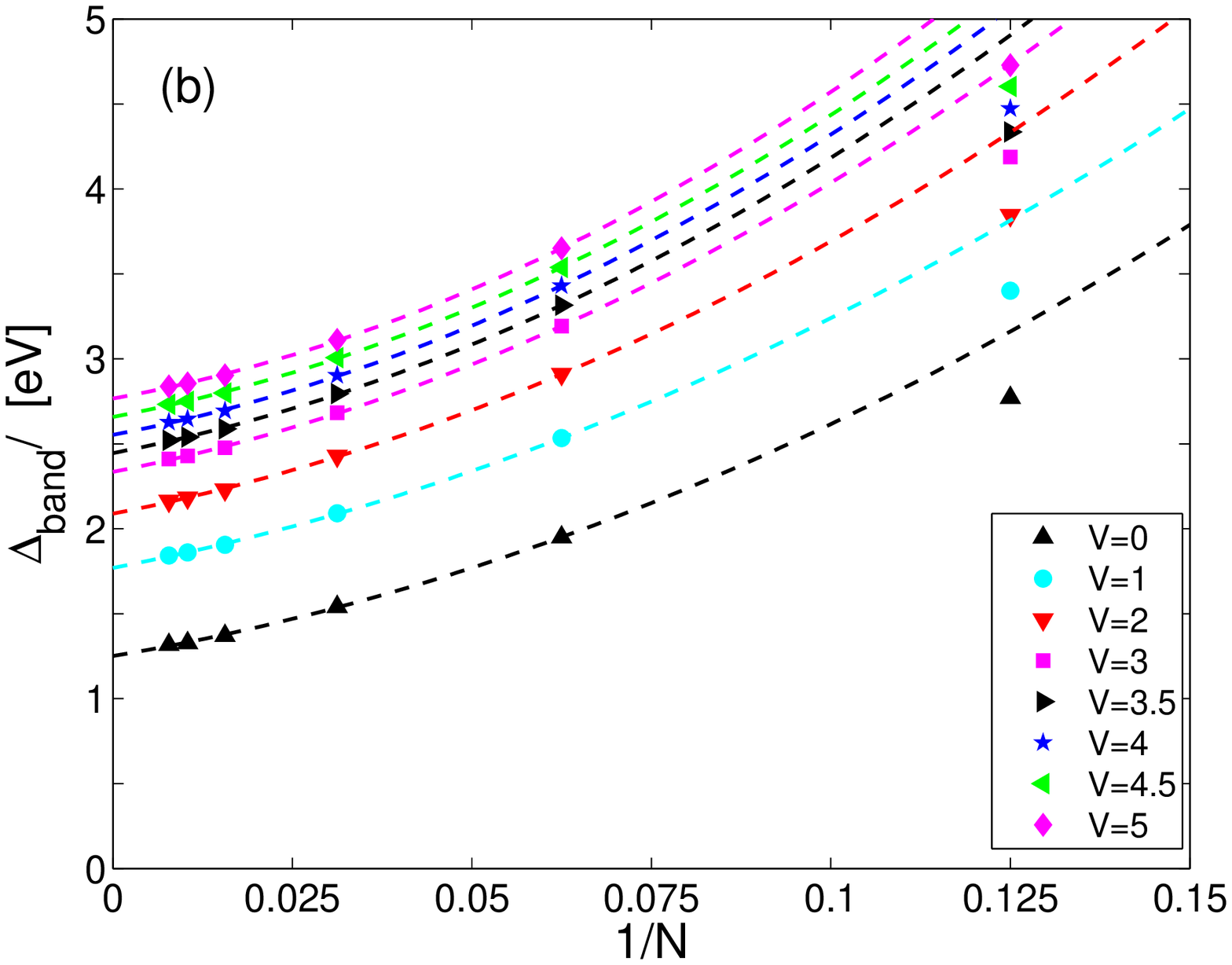}
\caption{(Color online) Band gap as a function of 
inverse system size~$1/N$ for various values of $V$
for (a) the Hubbard--Ohno potential~(\protect\ref{eqn:Ohnoforfigure})
and (b) the screened potential~(\protect\ref{eqn:Veffforfigure}).
The lines are quadratic fits.\label{Fig:gap-ohno}}
\end{center}
\end{figure}

When we target several eigenstates simultaneously, our
calculations on longer chains give less reliable results.
This limits the accuracy
of the results obtained from finite-size extrapolations. 
In such cases, we restrict our extrapolations to use
data up to $N_{\rm max}=80$.

\section{Results}
\label{sec:results}

\subsection{Single-particle gap, optical gap, and exciton binding energy}
\label{subsec:resultenergy}

In Fig.~\ref{Fig:gap-ohno} 
we show the single-particle gap as a function of inverse system size~$1/N$
for the Hubbard--Ohno potential~(\ref{eq:ohno-hubbard}) 
and the screened potential~(\ref{eq:erf}).
The lines are quadratic fits in the inverse system size.
The finite-size corrections to the result in the thermodynamic
limit, $N\to \infty$, are less than $0.05\, {\rm eV}$
for $N\gtrsim 100$.

As expected and as seen in Fig.~\ref{Fig:gap-ohno}, 
the single-particle gap increases
as a function of the Coulomb parameter~$V$.
For the chosen band-structure 
parameters~(\ref{eq:bsparameters}), the bare band gap is 
$\Delta(V=0)=1.25\, {\rm eV}$, which is only half as large as the observed
single-particle gap in PDAs. The Coulomb interaction
accounts for the other half of the single-particle gap, i.e.,
exchange and correlations play an important role in this class of materials.
In order to fit the experimentally observed gap, 
$\Delta_{\rm exp}=2.4\ {\rm eV}$, we choose $V_{\rm HO}=4.0\, {\rm eV}$
for the Hubbard--Ohno potential and $V_{\rm scr}=3.5\, {\rm eV}$
for the screened potential. These values correspond to a screening length
of $R_{\rm HO}=3.6\, \mbox{\AA}$ and $R_{\rm scr}=4.1\, \mbox{\AA}$,
respectively.

\begin{figure}[htb]
\begin{center}
\includegraphics[width=7.9cm]{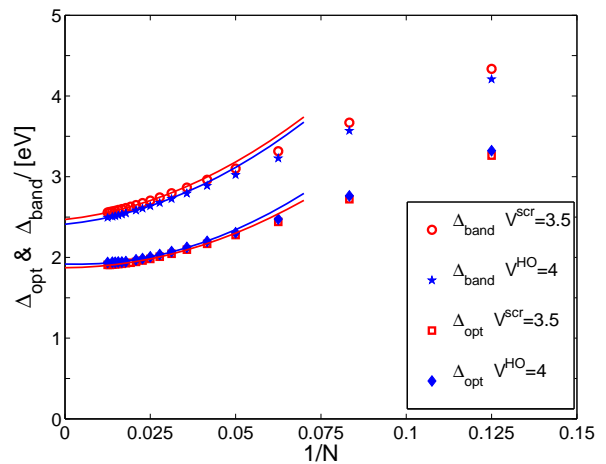}
\end{center}
\caption{(Color online) Band gap and optical gap as a function of 
inverse system size~$1/N$ for $V_{\rm HO}=4.0\, {\rm eV}$
for the Hubbard--Ohno potential~(\protect\ref{eq:ohno-hubbard})
and for $V^{\rm scr}=3.5\, {\rm eV}$
for the screened potential~(\protect\ref{eq:erf}).
The lines are fits to a quadratic polynomial in $1/N$.\label{Fig:optical-gap}}
\end{figure}

Both potentials display bound exciton states below the single-particle gap.
In Fig.~\ref{Fig:optical-gap} 
we show the single-particle gap $\Delta(N)$ and the
optical gap $\Delta_{\rm opt}(N)$ 
as a function of inverse system size~$1/N$ for both potentials.
Figure~\ref{Fig:optical-gap} shows that both effective potentials reproduce
the exciton energy, $\Delta_{\rm opt}\approx 1.9\, {\rm eV}$, for
PDA-DCHD.
Correspondingly, we can reproduce the experimentally observed 
exciton binding energy, $\Delta_{\rm bind}=0.5\, {\rm eV}$,
with both potentials.

\subsection{Oscillator strengths, dark states, and second exciton state}

\begin{figure}[htb]
\begin{center}
\includegraphics[width=8cm]{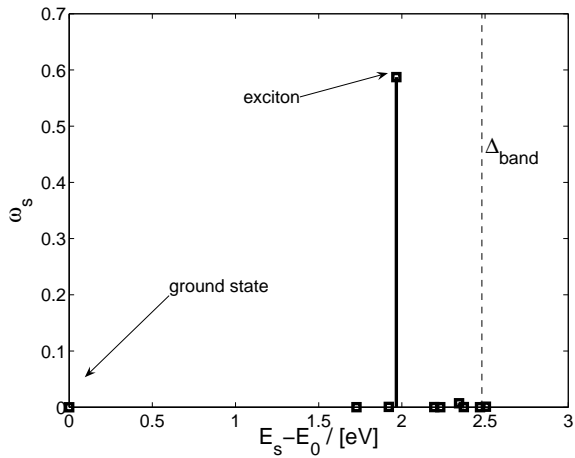}
\caption{Oscillator strengths $w_{\rm s}$, Eq.~(\protect\ref{eq:weights}),
as a function of energy for the Hubbard--Ohno potential 
($V_{\rm HO}=4.0\, {\rm eV}$) for $N=80$. The weight of the primary
exciton is $w_3=0.59$; 
the total weight of the first nine optically excited states is 
$\sum_{i=1}^9 w_i=0.60$.\label{Fig:oszi-strength}}
\end{center}
\end{figure}

We show the distribution of oscillator strengths
for the Hubbard--Ohno potential
in Fig.~\ref{Fig:oszi-strength}. The screened potential 
with $V_{\rm scr}=3.5\, {\rm eV}$ leads to 
a qualitatively similar distribution in that the majority of the weight lies
in the primary exciton.
For $N=80$ sites, the oscillator strength
for the primary exciton is $w_{\rm ex, HO}=0.59$
for the Hubbard--Ohno potential for $V_{\rm HO}=4.0\, {\rm eV}$.
The excitons carry about 60\% of the total spectral weight.
This is in good agreement with experiment
where the spectral weight of about $n_{\rm eff}=1.2$
of $n_{\rm tot}=2$ $\pi$-electrons is found below the
single-particle gap; see Fig.~5 of Ref.~\onlinecite{Weiser}.

We note that there are two optically dark states
{\sl below\/} the primary exciton, with
energy differences 
$\Delta_{A1}=0.24\, {\rm eV}$ and $\Delta_{A2}=0.05\, {\rm eV}$.
Both states are spin triplets.
Experimentally, such dark states have been located at 
$\Delta_{A}=0.4\, {\rm eV}$.~\cite{Lawrenceetal}
Therefore, we obtain a qualitatively and even semi-quantitatively
correct ordering of the excited states.

The primary exciton at the optical excitation energy 
$\Delta_{\rm opt}(N=80)=1.97\, {\rm eV}$ carries 99\% of the excitonic
weight. 
The second exciton around 
$\Delta_{\rm opt}'(N=80)=2.34\, {\rm eV}$ carries only 1\% of the excitonic
weight. Thus, our calculations indicate that two excitons should
be visible in the PDA chains, whereby the second exciton has a binding energy
of $\Delta_{\rm bind}'=0.1\, {\rm eV}$ and is lower in intensity by two orders 
of magnitude. Experimentally, it is difficult to detect
the second exciton because it is hidden by the intense phonon replicas
of the primary exciton.

\subsection{Exciton wave function and exciton radius}
\label{sect:singletwavefunction}

\begin{figure}[htb]
\begin{center}
\includegraphics[width=8cm]{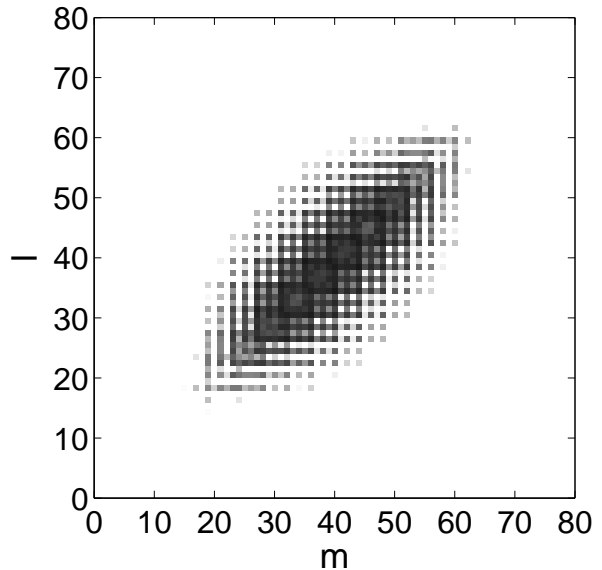}
\caption{Probability distribution 
$P_{\rm ex}(l,m)$~(\protect\ref{Pij})  
to find a hole at site~$l$ 
and an electron at site~$m$ in the exciton state. 
This `exciton wave function' is shown
for the Hubbard--Ohno potential with $V_{\rm HO}=4.0\, {\rm eV}$ 
on a chain with $N=80$ sites.\label{Fig:wavefunction-hubbard-ohno}}
\end{center}
\end{figure}

In the following, we concentrate on the primary exciton.
In Fig.~\ref{Fig:wavefunction-hubbard-ohno}
we show the probability distribution $P_{\rm ex}(l,m)$, Eq.~(\ref{Pij}),
i.e., the `exciton wave function' for the Hubbard--Ohno potential
for $V_{\rm HO}=4.0\, {\rm eV}$. The probability distribution is similar
for the screened potential for $V_{\rm scr}=3.5\, {\rm eV}$.

The probability distribution reflects the structure 
of the unit cell with four carbon atoms.
We expect that the exciton wave function factorizes,
$P_{\rm ex}(l,m)\approx \Psi_{\rm CM}[(l+m)/2]\varphi(|l-m|)$,
where $\Psi_{\rm CM}[(l+m)/2]$ describes the motion of the
center-of-mass and $\varphi(|l-m|)$ describes its internal structure.
The center-of-mass wave function follows that of a particle in a box. 
In an infinite system, it corresponds to
a state with zero total momentum because the light field
adds only a negligible momentum to the ground state.
In fact, for fixed $|l-m|$, we observe
nodes at the boundaries ($l,m\to 0,N$) and a maximum in the middle
of the chain for $l,m\approx N/2$.
For fixed center-of-mass coordinate $l+m$, the probability distribution
$P_{\rm ex}(l,m)$ reveals the internal structure of the exciton,
$\varphi(|l-m|)$. Cross sections of $P_{\rm ex}(l,m)$ 
along the lines $l+m={\rm const.}$
show that electron and hole are bound to each other, 
i.e., $\varphi(|l-m|)$ is vanishingly small for $|l-m|> r_{\max}$.

The exciton wave function $P_{\rm ex}(l,m)$
shows a prominent odd-even effect as a function
of the particle-hole separation $|l-m|$.
This is a consequence 
of the invariance of the Hamiltonian and of the current operator
under a particle-hole transformation, see Sect.~\ref{subsec:phtransform}.
At half band-filling, 
the ground state $|\Phi_0\rangle$
is invariant under this transformation. If the same applies
to an excited state $|\Phi_s\rangle$,
an inversion symmetric system obeys $\langle \Phi_s |\hat{c}_{l,\sigma}^+
\hat{c}_{m,\sigma}|\Phi_0\rangle
=(-1)^{l+m+1}\langle \Phi_s |\hat{c}_{l,\sigma}^+
\hat{c}_{m,\sigma}|\Phi_0\rangle$. Therefore,
the overlap vanishes for even $|l-m|$. Since 
the exciton obeys $|\Phi_{\rm ex}\rangle =
\sqrt{w_{\rm ex}}\hat{J}|\Phi_0\rangle +|\Phi_{\rm rest}\rangle$,
and the system is approximately inversion symmetric,
it is only the (small) contribution $|\Phi_{\rm rest}\rangle$
which contributes to the exciton wave function for even $|l-m|$.

\begin{figure}[htb]
\begin{center}
\includegraphics[width=8cm]{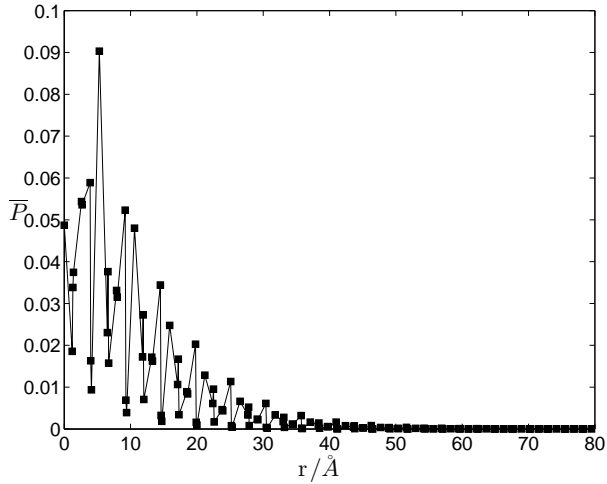}
\caption{Radial distribution function
$\overline{P}_{\rm ex}(r)$~(\protect\ref{pbar})
for the Hubbard--Ohno potential ($V_{\rm HO}=4.0\, {\rm eV}$) 
for a chain of $N=80$ sites.\label{Fig:radial-ohno-hubbard}}
\end{center}
\end{figure}

Further insight into the structure of the exciton wave function
is gained from the radial distribution function
$\overline{P}_{\rm ex}(r)$, Eq.~(\ref{pbar}),
for the electron-hole distance, which 
is shown in Fig.~\ref{Fig:radial-ohno-hubbard} for a chain 
of $N=80$ sites.
As expected for a bound electron-hole pair,
the distribution decays rapidly as a function of 
the electron-hole distance~$r$.
The peak at a distance $r\approx 4$ reflects the 
fact that there are four carbon atoms in the unit cell.
The oscillations in the radial probability are the result of the odd-even effect
observed in the probability distribution function 
$P_{\rm ex}(l,m)$, Eq.~(\ref{Pij}).
The overall behavior of the radial distribution resembles
the results obtained from the $G$$W$+BSE approach to 
polymers.~\cite{LouieRohlfing,LDA-GW-BSE,van-der-Horst}

\begin{figure}[htb]
\begin{center}
\includegraphics[width=8cm]{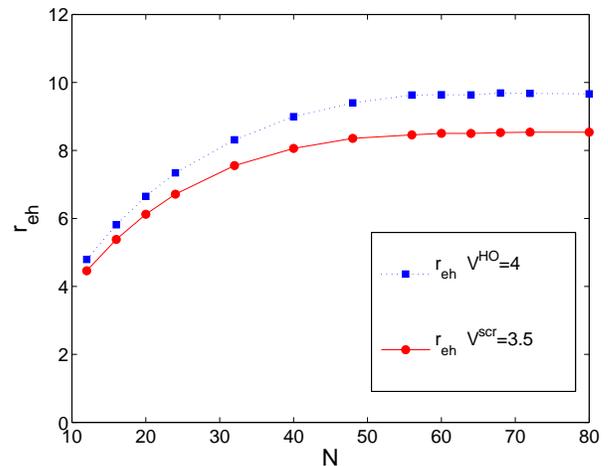}
\caption{Exciton radius as a function of system size
for the Hubbard--Ohno potential 
($V_{\rm HO}=4.0\, {\rm eV}$) and 
the screened potential ($V_{\rm scr}=3.5\, {\rm eV}$).
\label{Fig:radius-ohno-hubbard}}
\end{center}
\end{figure}

Finally, we show the exciton radius, Eq.~(\ref{<r>}),
as a function of system size for the Hubbard--Ohno potential
for $V_{\rm HO}=4.0 \, {\rm eV}$ and the screened potential
for $V_{\rm scr}=3.5\, {\rm eV}$ in Fig.~\ref{Fig:radius-ohno-hubbard}.
As expected for a bound exciton, it does not increase much with system
size for $N\gtrsim 40$. 
We find an extrapolated exciton radius $r_{\rm eh}^{\rm HO}=9.67\, \mbox{\AA}$
and $r_{\rm eh}^{\rm scr}=8.54\, \mbox{\AA}$
for the Hubbard--Ohno and screened potentials, respectively.

\begin{figure}[htb]
\begin{center}
\includegraphics[width=8cm]{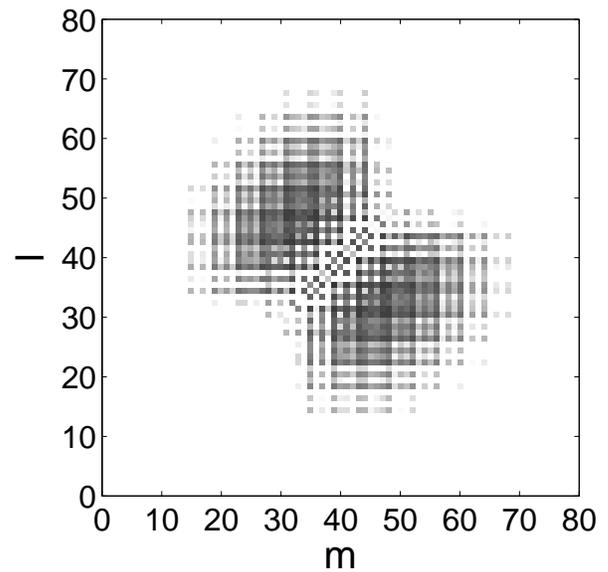}
\caption{Probability distribution 
$P_{\rm ex}(l,m)$~(\protect\ref{Pij})  
for the second exciton for the Hubbard--Ohno potential 
with $V_{\rm HO}=4.0\, {\rm eV}$ 
on a chain with $N=80$ sites.\label{Fig:wavefunction-hubbard-ohno-2nd}}
\end{center}
\end{figure}

For completeness, in Fig.~\ref{Fig:wavefunction-hubbard-ohno-2nd},
we show the wave function of the second, weak singlet exciton.
The center-of-mass coordinate again describes a particle in the box,
whereas the internal structure
for the relative motion of electron and hole displays a node 
as a function of $|l-m|$ for $l+m=$const. 
The size of the second exciton is about a factor of two larger than
the size of the primary exciton.

\subsection{Polarizability}
\label{sect:polari}

The polarizability follows from the Stark shift of the exciton in the
presence of an external electric field.
When we measure the strength of the electrical field~$F$
in terms of the energy unit $f=Fea_0$, with $a_0=1\, \mbox{\AA}$,
we can write the polarizability in the form
\begin{equation}
\frac{p}{{a_0}^3} = 28.694 \frac{\lambda}{({\rm eV})^{-1}}\; , 
\end{equation}
where $\lambda$ describes the excitonic Stark shift,
$\delta \Delta_{\rm opt} = -\lambda f^2$, see Eq.~(\ref{eq:Starkshift}).

\begin{figure}[htb]
\begin{center}
\includegraphics[width=8cm]{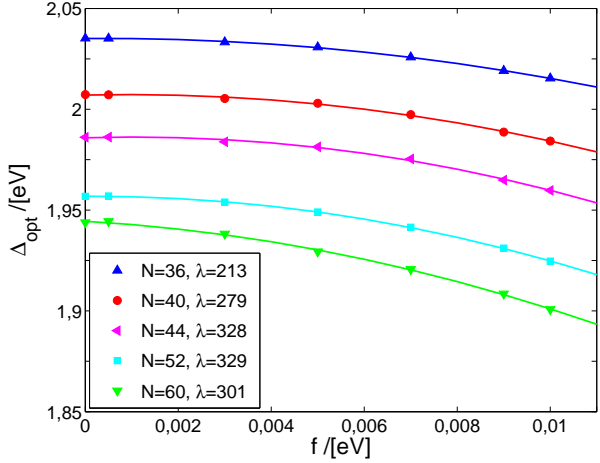}
\caption{(Color online) Stark shift of the exciton binding energy
as a function of the rescaled electrical field~$f=Fea_0$ ($a_0=1\, \mbox{\AA}$) 
for various system sizes for the Hubbard--Ohno potential 
($V_{\rm HO}=4.0\, {\rm eV}$).\label{Fig:Starkshift}}
\end{center}
\end{figure}

Fig.~\ref{Fig:Starkshift} shows the redshift of the binding energy
for the primary exciton
due to the electric field for various system sizes. For systems $N>40$,
the curvature $\lambda$ does not change significantly. This reflects
the fact that the exciton wave function does not depend on 
the system size for $N\gtrsim 40$.
Taking a rough typical value from the fits in Fig.~\ref{Fig:Starkshift},
we estimate
$\lambda\approx 3.0\cdot 10^2\, ({\rm eV})^{-1}$ 
so that we find $p=8.6\cdot 10^3 \mbox{\AA}^3$
for the polarizability. This compares favorably with the
experimental value for PDA-DCHD, $p_{\rm DCHD}=8.2 \cdot 10^3 \mbox{\AA}^3$,
or PDA-PTS, $p_{\rm PTS}=7.2 \cdot 10^3 \mbox{\AA}^3$.~\cite{Weiser}

In the experimental work,\cite{Weiser} a semi-empirical model
was used to extract the exciton radius~$r_{\rm exc}$ from the polarizability,
$p=\alpha (er_{\rm exc})^2/\Delta_{\rm opt}$, where $\alpha$
is a factor of the order of unity. For $\alpha=1$, the experimental value
for $p$ leads to $r_{\rm exc}\approx 12\,\mbox{\AA}$,
in good agreement with our value for the average
electron-hole separation, $r_{\rm eh}=9.7\,\mbox{\AA}$.

\subsection{Triplet exciton}
\label{sec:tripletsector}

\begin{figure}[htb]
\begin{center}
\includegraphics[width=7.9cm]{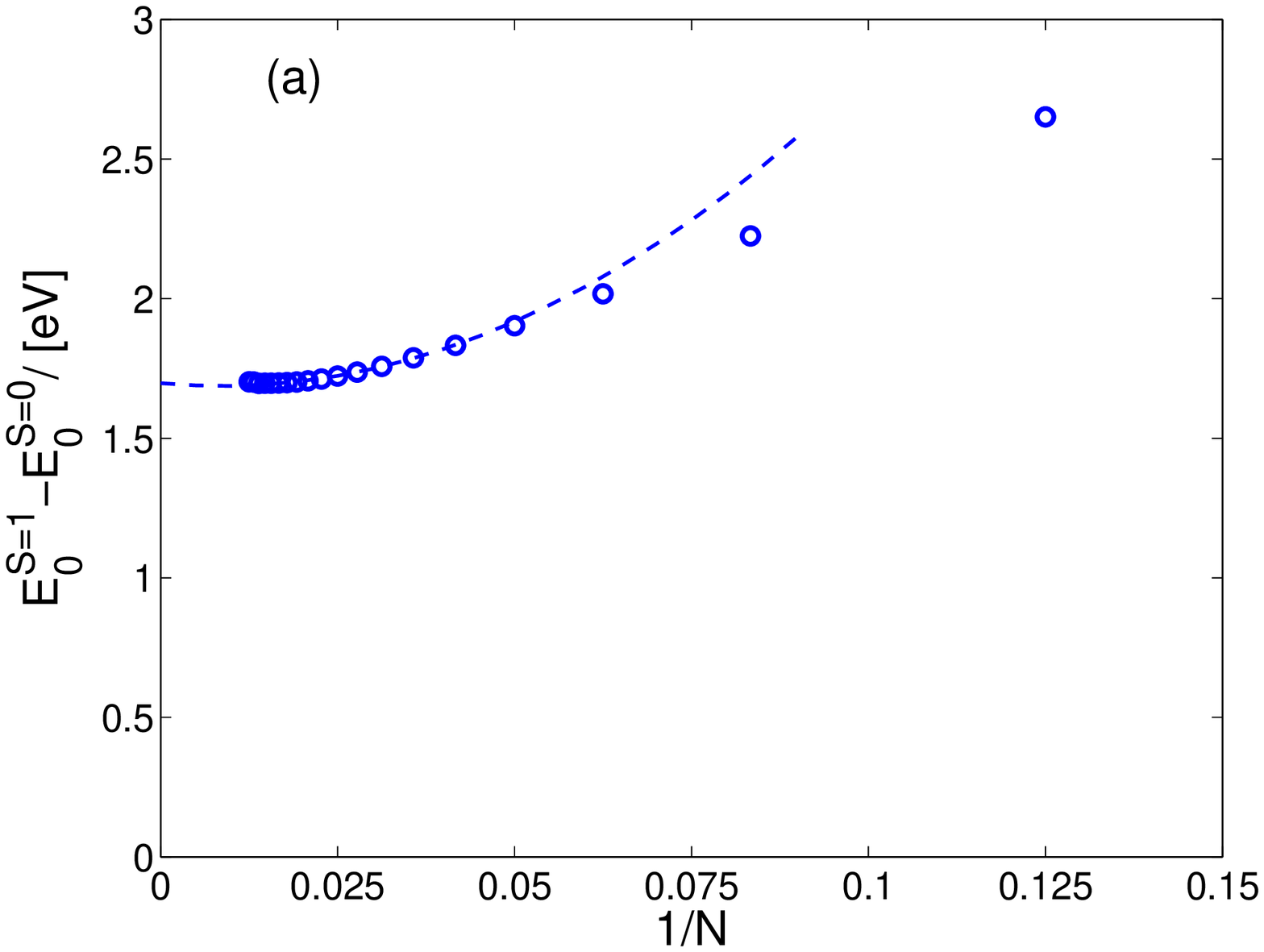} \\
\includegraphics[width=7.9cm]{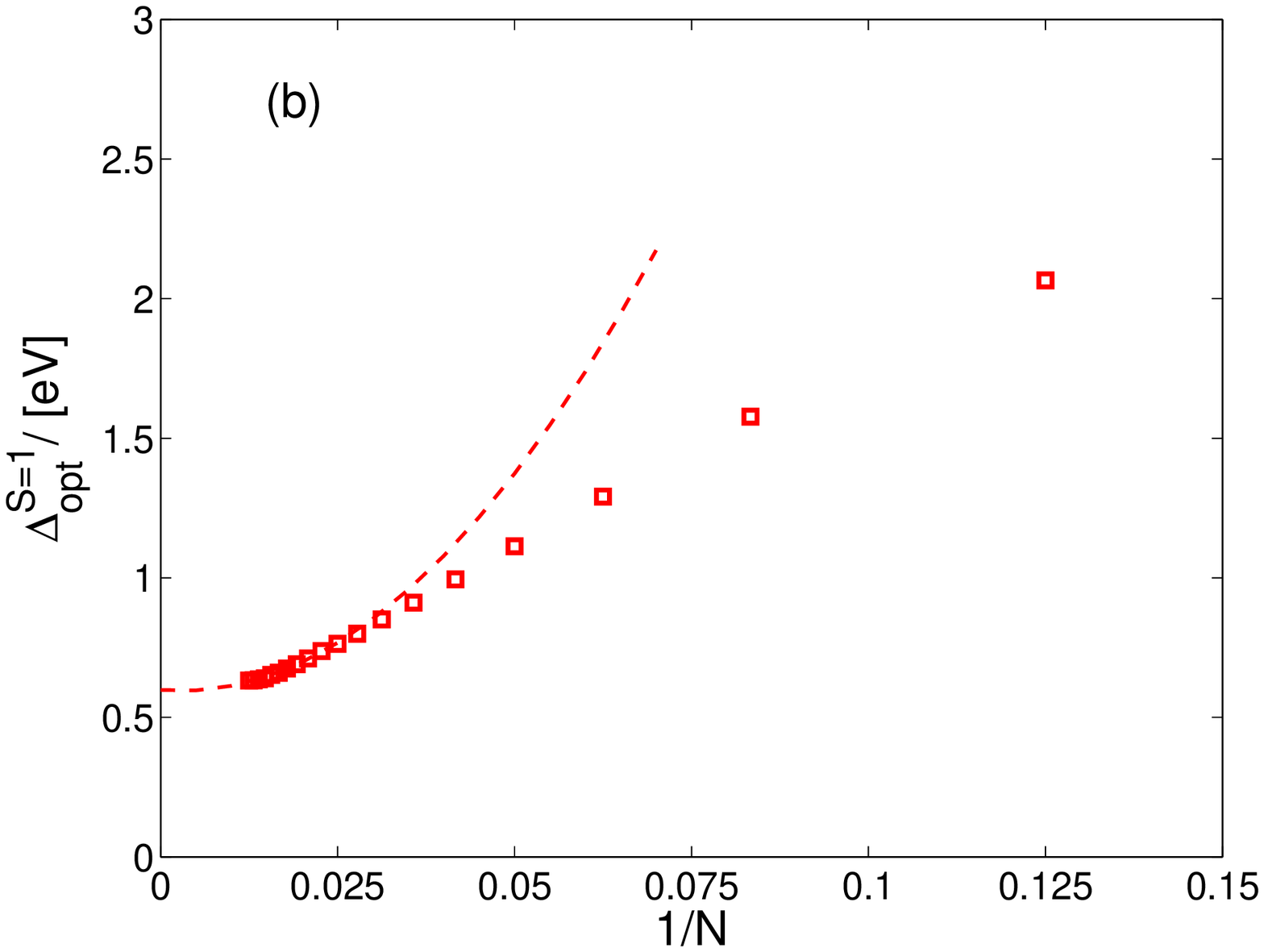}
\caption{(a) Energy difference between
the ground states of the two spin sectors
and (b) the energy of the triplet exciton as a a function
of inverse system size $1/N$ for the Hubbard--Ohno potential
($V_{\rm HO}=4.0\, {\rm eV}$).\label{energies-triplet}}
\end{center}
\end{figure}

Finally, we summarize our results for the triplet sector, i.e., excitations
with total spin $S=1$. Note that it is difficult to access this 
spin sector experimentally. In Fig.~\ref{energies-triplet},
we show the differences in the ground state energies in the spin singlet
sector ($S=0)$ and the spin triplet sector ($S=1$), together
with the optical gap in the triplet sector as a function of inverse
system size for the Hubbard--Ohno potential 
($V_{\rm HO}=4.0\, {\rm eV}$). As can be seen, one finds two finite gaps 
with different sizes which we will interpret below.
The results for the screened
potential ($V_{\rm scr}=3.5\, {\rm eV}$) are very similar.

\begin{figure}[t]
\begin{center}
\includegraphics[width=8cm]{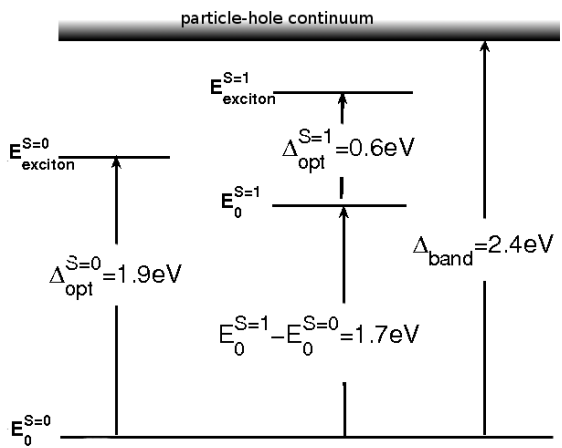}
\caption{Energy level scheme for the spin singlet and spin triplet
sector for the Hubbard--Ohno potential 
($V_{\rm HO}=4.0\, {\rm eV}$).\label{Fig:termscheme}}
\end{center}
\end{figure}

In Fig.~\ref{Fig:termscheme} we depict the resulting energy level scheme
for the Hubbard--Ohno potential for $V=4.0\, {\rm eV}$.
The energies are extrapolated values in the thermodynamic limit.
The ground state of the triplet sector is $E_0^{S=1}-E_0^{S=0}=1.7\, {\rm eV}$
higher in energy than the ground state for the singlet sector,
see Fig.~\ref{energies-triplet}a.
Therefore, this state is frequently called the `triplet exciton'.
This state is $0.2\, {\rm eV}$ below the singlet exciton.
Experimentally, however, this state has been detected at 
$0.9\, {\rm eV}$ below the singlet exciton.~\cite{tripletgap}
Therefore, our description underestimates this energy, i.e.,
the triplet ground state should be much lower in energy.
This discrepancy could be the consequence of
the large polaronic effects in the triplet sector, as has been
suggested in Ref.~\onlinecite{Barfordrelax}.
The dominant
optical excitation in the triplet sector 
lies only $0.6\, {\rm eV}$ above the
triplet ground state, see Fig.~\ref{energies-triplet}b, 
but still $0.4\, {\rm eV}$ above the
singlet exciton and is close to the single-particle gap, i.e., 
only $0.1\, {\rm eV}$ below the particle-hole continuum.

\begin{figure}[tb]
\begin{center}
\includegraphics[width=6.5cm]{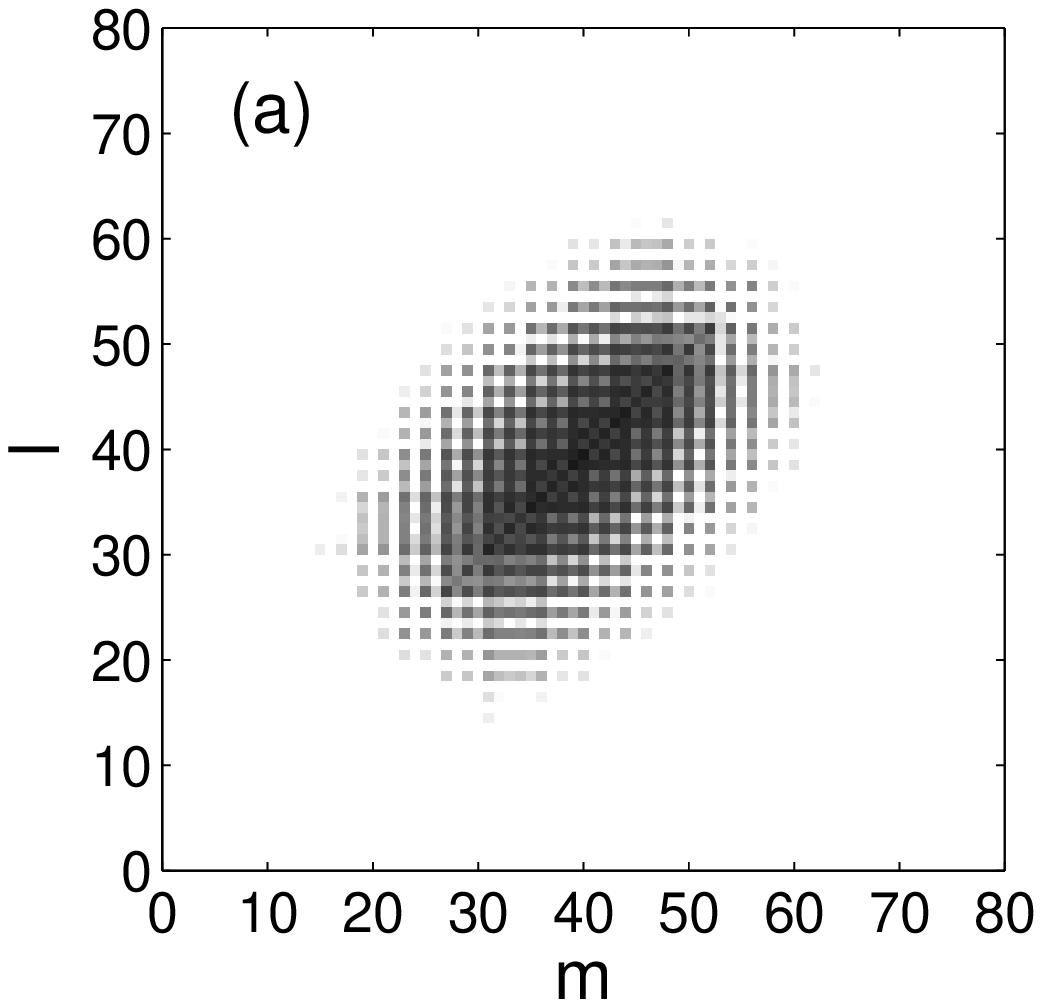}
\\
\includegraphics[width=6.5cm]{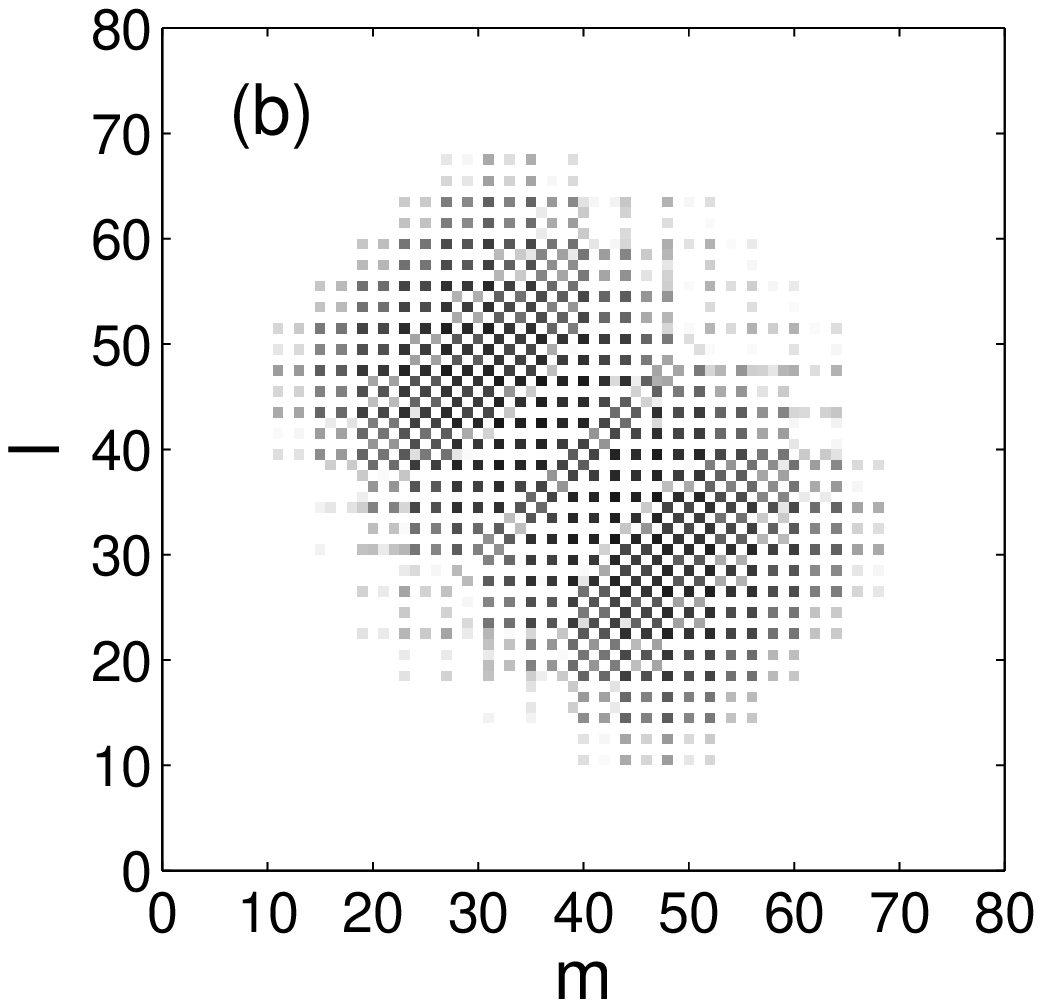}
\caption{Probability distributions 
$P_{\rm ex}(l,m)$, Eq.~(\protect\ref{Pij}),  
for (a) the triplet ground state and (b) its optical excitation
for the Hubbard--Ohno potential with $V_{\rm HO}=4.0\, {\rm eV}$ 
on a chain with $N=80$ sites.\label{Fig:tripletexciton}}
\end{center}
\end{figure}

The wave function of the triplet excitations is shown in 
Fig.~\ref{Fig:tripletexciton}.
Similarly to the singlet exciton discussed in
Sect.~\ref{sect:singletwavefunction}, 
the probability distribution 
approximately factorizes into a center-of-mass wave function
and a wave function for the relative coordinate. The
center-of-mass wave function are similar in all cases.
The wave function for the relative motion of electron and hole 
shows two nodes for the optical excitation
of the triplet ground state, whereas it has one node 
in the case of the optical excitation of the singlet ground state.

\section{Conclusions}
\label{sec:conclusions}

In this paper, we have studied the role of electron-electron 
interactions in poly-diacetylenes.
Since the bare band gap is only half as large as the observed single-particle 
gap and since the binding energy of the singlet exciton of 0.5~eV
is 20\% of the single-particle gap, exchange and correlation must play 
an important role in this class of materials. 
Our density-matrix renormalization group method permits
the numerically exact treatment of an appropriate model Hamiltonian 
with long-range Coulomb interactions for 
a large number of electrons.
We have used the experimentally observed single-particle gap,
$\Delta_{\rm band}\simeq 2.4\, {\rm eV}$, to fix the strength
of the Coulomb interaction at short distances. We have determined
the lowest-lying optical excitation and have reproduced the
observed exciton binding energy 
in poly-diacetylenes, $\Delta_{\rm bind}\simeq 0.5\, {\rm eV}$.

The key difference between our work and previous numerical
DMRG studies~\cite{Bursill,Barfordrelax}
is the parameterization of the Coulomb interaction.
We argue that the Ohno parameterization of the Pariser--Parr--Pople 
interaction~\cite{PPP}
is not appropriate for very short distances because
the local (Hubbard) interaction is too small.
We propose to use the Hubbard--Ohno potential~(\ref{eq:ohno-hubbard})
or the full expression~(\ref{eq:erf}) of the
Coulomb interaction for effectively one-dimensional structures.
Our findings support earlier theoretical studies of 
poly(para-phenylene vinylene).~\cite{ChandrossMazumdar}

Our results indicate that the PDA chain has two optically dark states
below the exciton, in qualitative agreement with experiment.
Moreover, the screening potential supports a second bound exciton 
with binding energy $\Delta_{\rm bind}'\simeq 0.1\, {\rm eV}$
whose intensity is two orders of magnitude smaller than
that of the primary exciton.

The exciton wave function approximately factorizes
into two terms: the center-of-mass wave function and the 
relative wave function. The former describes the excitonic 
`particle-in-a-box' state; the latter describes the pair state of an
electron and a hole whose separation (exciton radius) 
rapidly converges to a finite value with increasing system size.
In order to investigate the behavior of the exciton in the presence of an
external electric field, we have studied the polarizability due to 
the Stark shift of the exciton. As expected for a bound electron-hole pair,
the exciton displays a quadratic Stark redshift in energy as a function
of the field strength. Our calculated polarizability 
reproduces the experimental results for PDA-DCHD and PDA-PTS chains. 

Finally, we have studied the triplet sector.
The energy of the triplet ground state found in our calculation is too high,
i.e., the binding energy of the lowest-lying triplet excitation
is too small. We attribute this 
discrepancy to strong polaronic
effects in the triplet sector.~\cite{Barfordrelax}
In our work, we have not considered the effects of lattice relaxation,
the electrostatic potential of the poly-diacetylene side-groups,
and geometry effects. The inclusion of these effects is required
for a more detailed description of individual
members of the poly-diacetylene family.

\begin{center}
{\bf Acknowledgments}
\end{center}
We thank Gerhard Weiser, Michel Schott, and Walter Hoyer for useful discussions,
and J\"org Rissler for his important 
contributions at an early stage of this project.
This work was supported in part by the center {\sl Optodynamik\/} 
of the Philipps-Universit\"at Marburg,
by the Deutsche Forschungsgemeinschaft
(GE~746/7-1 and GRK~790), and by the Hungarian Research Fund (OTKA)
Grants Nos.\ K~68340 and K~73455. 
The authors acknowledge computational support from Dynaflex Ltd.\ 
under Grant No.\ IgB-32, and
thank the Erwin-Schr\"odinger Institute for Mathematical Physics 
for its hospitality where part of this work was accomplished.

\end{document}